\documentclass[12pt,a4paper]{article}
\usepackage{amssymb}
\usepackage{graphicx}
\usepackage{cite}

\begin{document}

\title{Theory of Rochelle salt: beyond the Mitsui model}

\author{I.V.~Stasyuk and O.V.~Velychko\\ [2ex]
Institute for Condensed Matter Physics\\
of National Academy of Sciences of Ukraine,\\
1 Svientsitsky Str., 79011 Lviv, Ukraine
}

\maketitle

\centerline{%
\parbox{0.9\textwidth}{%
A simple four-sublattice order-disorder model is developed for
description of phase transitions and dielectric properties of the
Rochelle salt crystal. The model is developed as a generalization
of the semimicroscopic Mitsui model. The symmetry properties of
lattice and spatial orientations of effective dipoles connected
with the asymmetric structure units in the elementary cell are
taken into account. The model allows to investigate the
temperature and field behaviour of transverse (besides
longitudinal) components of dielectric susceptibility. The
influence of the transverse electric field
$\vec{E}\parallel\vec{b}$ on the phase transition points and
spontaneous polarization is studied.
%64.60.Cn Order-disorder transformations; statistical mechanics of model
%77.22.-d Dielectric properties of solids and liquids
%77.80.-e Ferroelectricity and antiferroelectricity
%77.80.Bh Phase transitions and Curie point
%77.80.Fm Switching phenomena
%77.84.-s Dielectric, piezoelectric, ferroelectric, and antiferroelectric
%         materials (for nonlinear optical materials, see 42.70.Mp; for
%         dielectric materials in electrochemistry, see 82.45.Un)
}%
}

\section{Dielectric properties of Rochelle salt}

Rochelle salt (RS) is a particular object in the family of
ferroelectric crystals with hydrogen bonds. Despite the fact that
study of its properties has a long history, the structural aspects
and mechanisms of phase transitions in this crystal are not
conclusively established.
RS becomes ferroelectric (with spontaneous polarization parallel
to the crystallographic $a$-axis) in the narrow temperature range
between 255~K and 297~K.
Both nonpolar phases are orthorhombic ($P2_{1}2_{1}2_{1}$), while
the polar phase is monoclinic ($P2_{1}11$).
An
elementary cell consists of four formula units.

Numerous data of structural investigations (starting from the
early results obtained by the X-ray spectroscopy \cite{Bee41} and
neutron scattering \cite{Fra54}) do not give a definitive answer
to the question of microscopic nature of
phase transitions in RS.
Dielectric relaxation in the microwave frequency region and the
critical slowing down around the phase transitions
point to the order-disorder type scenario \cite{San68}.
Alternatively, the presence of the soft mode,
which was observed by the far infrared reflectivity and Raman
spectroscopy in the lower paraelectric phase \cite{Kam95} as well
as by microwave dielectric measurements\cite{Vol86}
is rather a manifestation of the displacive-type transition.

The soft mode in paraelectric phase is connected with structure
changes (such as displacement of the O(8) oxygen along the
$a$-axis, rotation of tightly coupled water molecules with O(9)
and O(10) ions) which take place at transition to the
ferroelectric phase \cite{Shi98}; it is confirmed by the inelastic
neutron scattering data \cite{Hli01}. Respective static
displacements are the reason of appearance of additional
dipole moments of local structure units at phase
transition to the ferroelectric phase.

Such displacements can be interpreted also as changes in the
population ratio of two sites in the disordered paraelectric
structure (revealed in the structure investigations
\cite{Shi01,Nod00}). Large values of the anisotropic temperature
factors can be also connected with local disorder
\cite{Suz94}. The existence of the double atomic positions was
taken into account in the so-called split-atom model for RS
\cite{Iwa89b}.

The order-disorder picture of phase transitions in RS forms the
basis of the semimicroscopic Mitsui model \cite{Mit58}, where
asymmetry of occupancy of double local atomic positions as well as
compensation of electric dipole moments induced in paraphase were
taken into account. In spite of simplicity (consideration was
restricted to two sublattices and induced local dipole
moments were described by means of pseudospins, $S^z=\pm 1/2$),
the model explains quite successfully appearance of two Curie
points and effect of deuteration \cite{Mit58,Vak73eng}.

In \cite{Lev03} the model was extended due to inclusion
of piezoelectric coupling to the external field.
One should also mention the phenomenological
Landau theory \cite{Koz88}, adapted for systems with a double
critical point, which is applicable to the RS crystal in a broad
range of pressure, substitution concentration of ammonium and
temperature.

\begin{figure}[!b]
\begin{center}
\tabcolsep=3pt%
\begin{tabular}{ccc}
\includegraphics[width=0.47\textwidth]{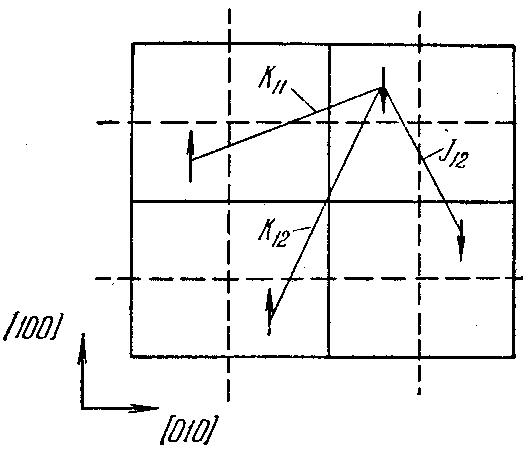}
&&
\includegraphics[width=0.47\textwidth]{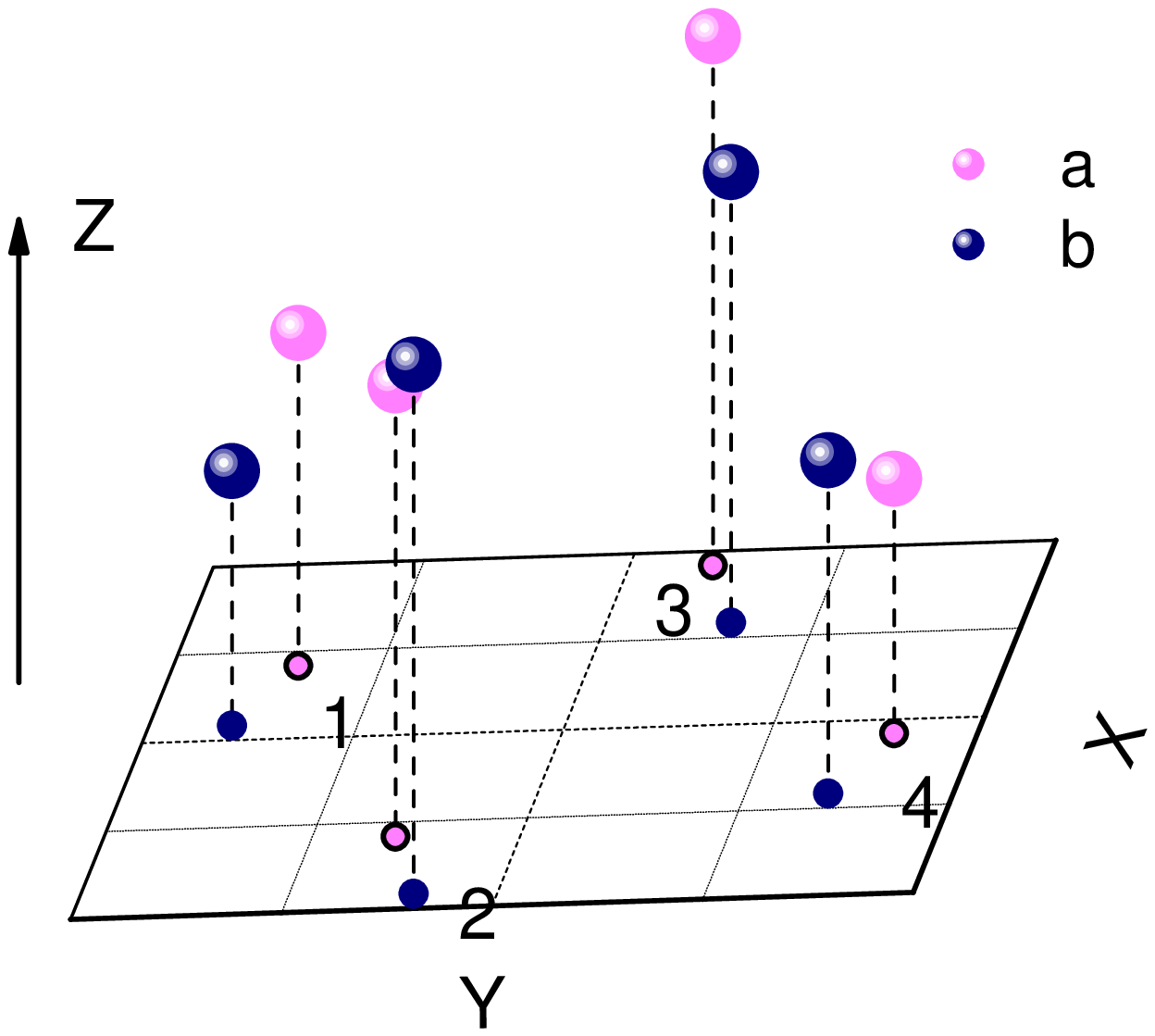}
\end{tabular}
\end{center}
\caption{Orientantions of dipole moments (ordering structure
units), producing resulting polarization, in the elementary cell
of the RS crystal: a comparison between the classical Mitsui
picture \protect{\cite{Vak73eng}} (left) and the proposed approach
(right). In the pseudospin formalism $\langle S^z \rangle =
\frac{1}{2}(n_a-n_b)$, where $n_{a,b}$ is a probability
(occupation) of respective orientation.}
\label{fig01}
\end{figure}

The Mitsui model simplifies real structure of the crystal a priori
choosing the ferroelectric axis among three two-fold axes thus
making an approach essentially ``one-dimensional''. It is
obviously insufficient for more complete description of dielectric
properties of the RS crystal. We can carry out a generalization,
making the model ``three-dimensional'' and taking into account the
presence of four (rather than two) translationally nonequivalent
groups of atoms in the unit cell (their positions are mutually
connected by elements of the point group of the crystal in
paraphase \cite{Bee41,Fra54}). Such structure units are
noncentrosymmetric. Effective dipole moments $\vec{\mu}_i$
$(i=1,\ldots,4)$ can be assigned to them as a whole; the sum of
these moments is equal to zero in paraphase.
Changes $\Delta\vec{\mu}_i$ in such dipole moments are responsible
for appearance of spontaneous polarization in the ferroelectric
state. Vectors $\Delta\vec{\mu}_i$ are oriented at the certain
angles to crystallographic axes and possess both longitudinal and
transverse components with respect to the $a$-axis
(Fig.~\ref{fig01}).

Let us use the order-disorder picture for description of such
changes. Taking into account double equilibrium positions of atoms
we come to the effective four-sublattice pseudospin model. The
model allows to calculate dielectric characteristics in any
direction and to consider also the effects caused by the influence
of the transverse electric field (applied perpendicularly to the
ferroelectric $a$-axis).

In the next section we propose a Hamiltonian obeying symmetry
properties of the crystal and derive expressions for main
thermodynamic characteristics of RS in the mean field
approximation. The obtained results of consideration of the
transverse field effect on the polarization and susceptibility are
presented in the third section.

\begin{figure}[!b]
\begin{center}
\begin{tabular}{ccc}
\raisebox{0.03\height}{\includegraphics[width=0.395\textwidth]{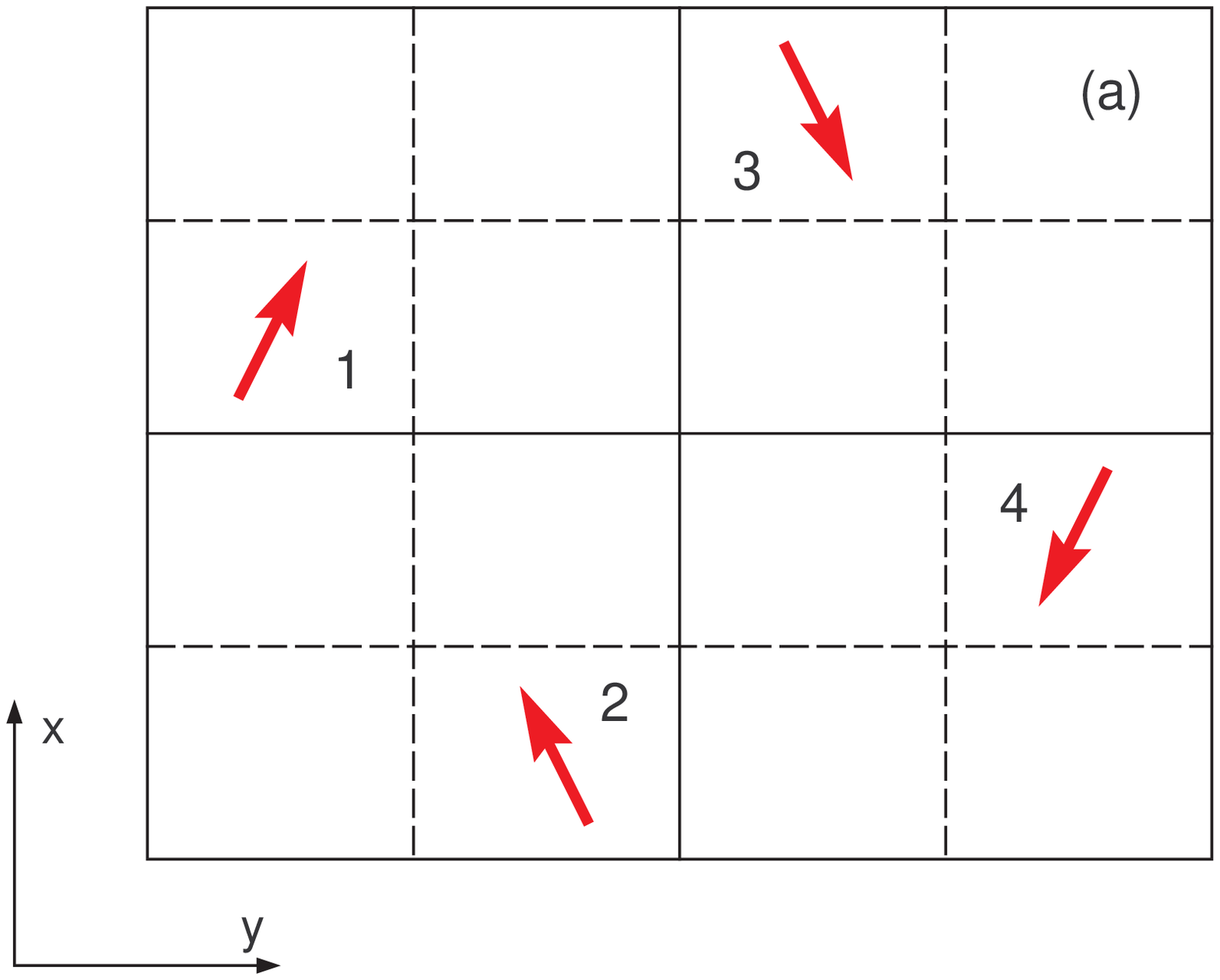}}
&&
\includegraphics[width=0.4\textwidth]{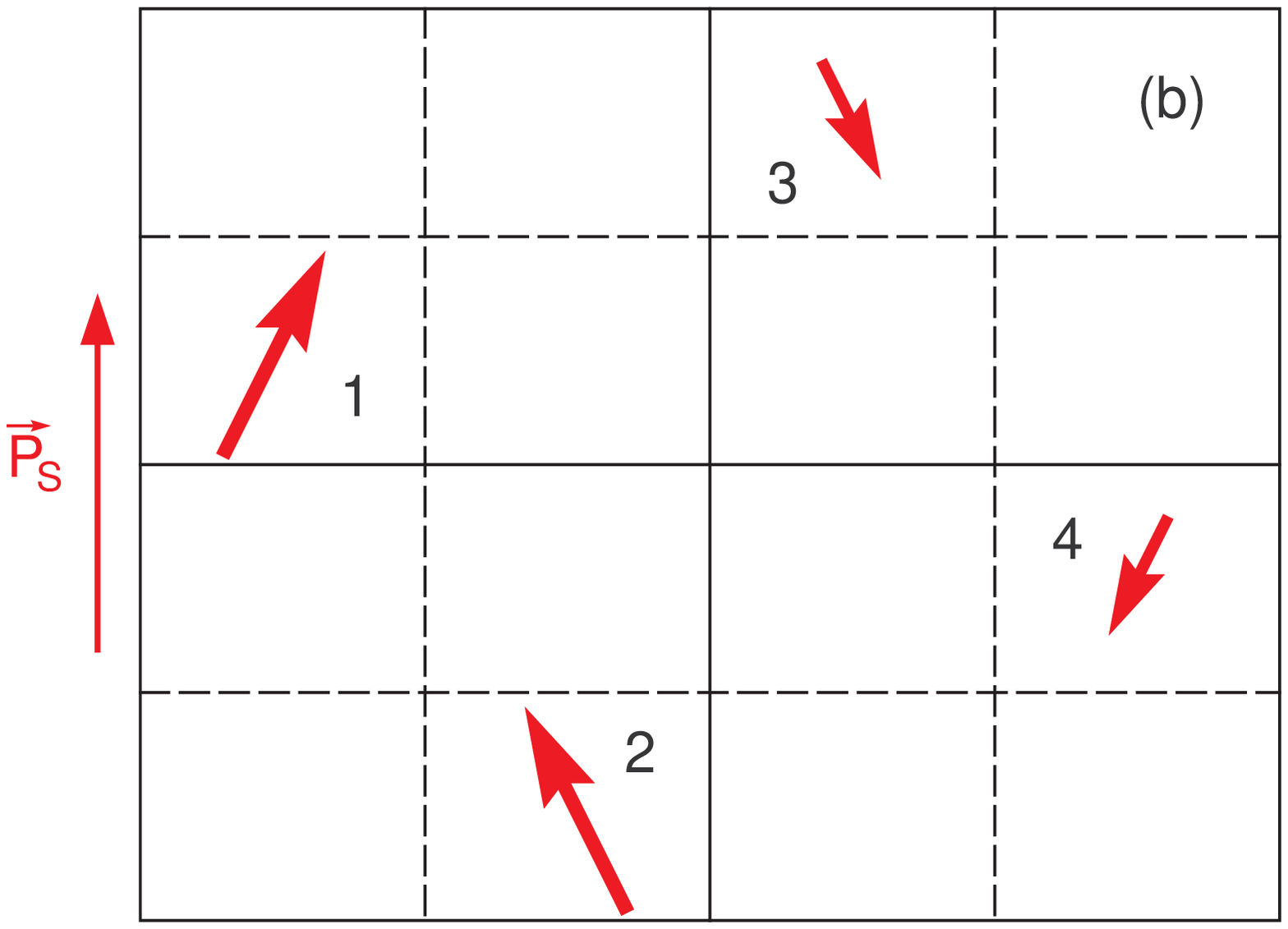}
\\
\strut &&  \\
\includegraphics[width=0.4\textwidth]{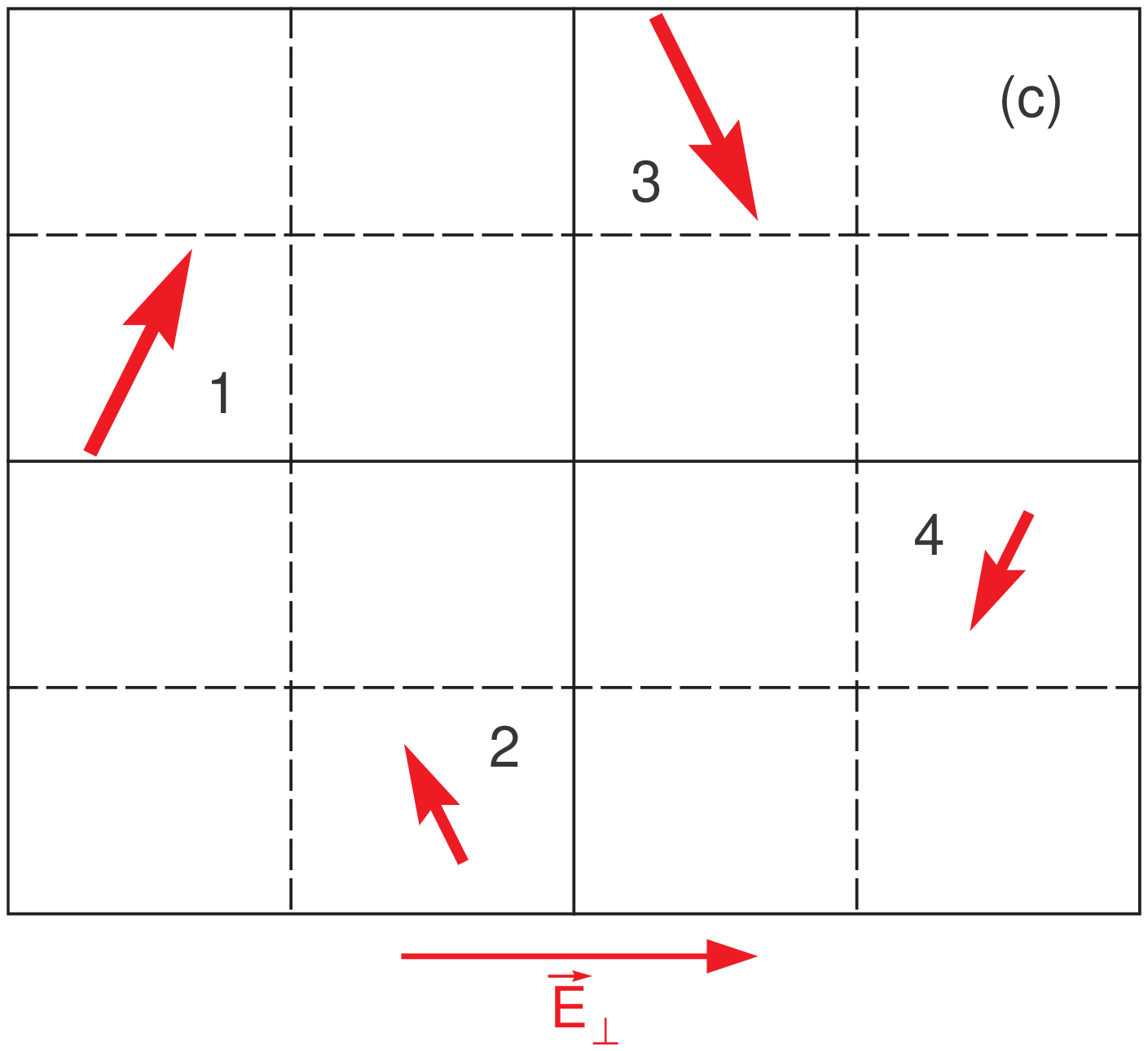}
&&
\includegraphics[width=0.4\textwidth]{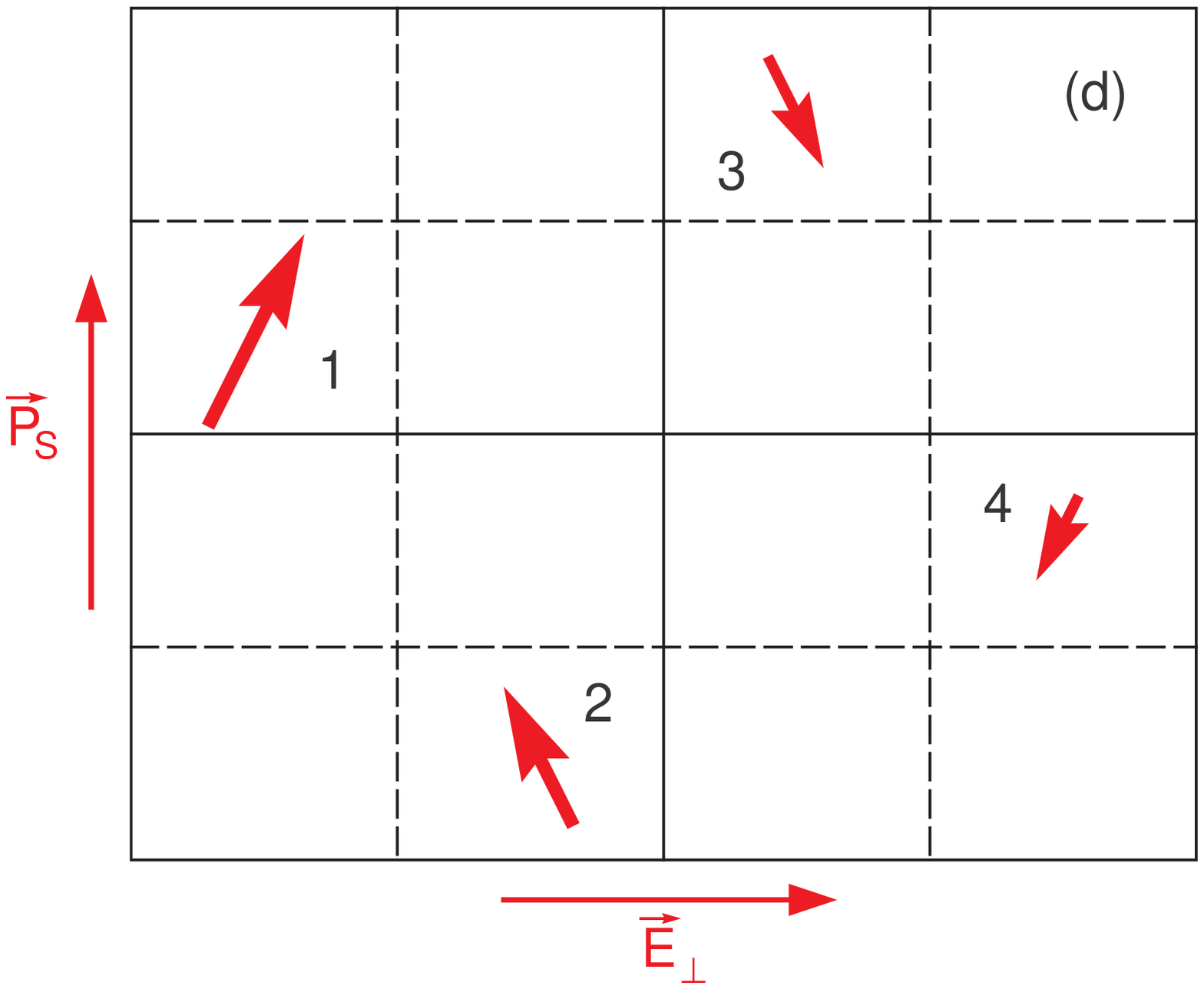}
\end{tabular}
\end{center}
\caption{Orientations of vectors $\Delta\vec{\mu}_{ik}$
($k=1,\ldots,4$) in a unit cell of the RS crystal and some
possible dipole orderings (projection on the plane $XY$):
(a) high symmetry phase (paraphase) -- absolute values of
pseudospins are equal in all sublattices;
(b) ferroelectric phase with $\vec{P}_{\mathrm{S}}\parallel X$ --
pseudospin values in sublattices 1 and 2 are larger;
(c) effect of the transverse field -- pseudospin values in
sublattices 1 and 3 are larger;
(d) $\vec{P}_{\mathrm{S}}\parallel X$ in the transverse field --
all values are different, sublattices 1 and 2 still prevail.%
}
\label{fig02}
\end{figure}

\section{Four-sublattice model: Hamiltonian and thermodynamics}

According to the above given arguments we take the four-sublattice
model as a base for simplified description of phase transitions
and dielectric properties of the RS crystal. Pseudospin variables
$S^z_{i1},\ldots,S^z_{i4}$ describe the before-mentioned changes
due to reordering of dipole moments of structure units:
$\Delta\vec{\mu}_{ik}\equiv \vec{d}_{k}S^z_{ik}$. Mean values
$\langle S^z \rangle = \frac{1}{2}(n_a-n_b)$ are related to
differences in position occupancies in the two-minima
representation of vectors $\Delta\vec{\mu}_{ik}$
(Fig.~\ref{fig02}).

We write down the Hamiltonian of the model in the pseudospin
representation:
\begin{eqnarray}
H &=&
-\frac{1}{2}\sum_{i \neq j}\sum_{k} J_{kk}(i,j) S^z_{ik}S^z_{jk}
-\frac{1}{2}\sum_{i,j}\sum_{k \neq l} K_{kl}(i,j) S^z_{ik}S^z_{jl}
\nonumber\\
&&{}
- \Delta \sum_{i} (S^z_{i1}+S^z_{i2}-S^z_{i3}-S^z_{i4})
- d_x E_x \sum_{ik} S^z_{ik}
\nonumber\\
&&{}
- d_y E_y \sum_{i} (S^z_{i1}-S^z_{i2}-S^z_{i3}+S^z_{i4})
- d_z E_z \sum_{i} (S^z_{i1}-S^z_{i2}+S^z_{i3}-S^z_{i4}),
\label{hmtn3d}
\end{eqnarray}%
where $J_{kk}(i,j)$ and $K_{kl}(i,j)$ describe the inter- and
intrasublattice interactions respectively. The internal field
$\Delta$ represents asymmetry in occupancies of the double
positions. The last three terms in Hamiltonian (\ref{hmtn3d})
describe an interaction with components $E_{\alpha}$
($\alpha=x,y,z$) of the external electric field.
For the sake of simplicity we do not include in (\ref{hmtn3d}) a
term describing tunnelling-like hoppings between equilibrium
positions.

Formula (\ref{hmtn3d}) can be considered as a generalization of
the Mitsui model Hamiltonian \cite{Mit58}: the first four terms
are similar to their analogs in that model. Besides the parameter
$\eta_1$, describing the ferroelectric ordering along the
$a$-axis, and the parameter $\xi$, responsible for the out of
phase ordering of the separated structure elements, there are new
parameters $\eta_2$ and $\eta_3$ related to dipole ordering along
the $b$- and $c$-axes, respectively:
\begin{eqnarray}
&&
\textstyle
\eta_1=\frac{1}{2}(\langle S^z_1 \rangle + \langle S^z_2 \rangle
                  +\langle S^z_3 \rangle + \langle S^z_4 \rangle),\qquad
\xi=\frac{1}{2}(\langle S^z_1 \rangle + \langle S^z_2 \rangle
               -\langle S^z_3 \rangle - \langle S^z_4 \rangle),
\nonumber\\ [1.5ex]
&&
\textstyle
\eta_2=\frac{1}{2}(\langle S^z_1 \rangle - \langle S^z_2 \rangle
                  -\langle S^z_3 \rangle + \langle S^z_4 \rangle),\qquad
\eta_3=\frac{1}{2}(\langle S^z_1 \rangle - \langle S^z_2 \rangle
                  +\langle S^z_3 \rangle - \langle S^z_4 \rangle).
\label{set1}
\end{eqnarray}

Considering Hamiltonian (\ref{hmtn3d}) in the mean field approximation
we obtain the following equations for average values of pseudospins
\begin{equation}
\langle S^z_k \rangle =
\textstyle
\frac{1}{2} \tanh \left(\frac{1}{2}\beta H_k \right),\qquad
k=1,\ldots,4.
\label{sz}
\end{equation}
Self-consisted internal fields $H_k$ are given by the expressions
\begin{eqnarray}
H_{1,2} &=&
\textstyle
(h_x+\frac{1}{2}\eta_1)
+(h-\frac{1}{2} a_1 \xi)
\pm(h_y-\frac{1}{2} a_2 \eta_2)
\pm(h_z+\frac{1}{2} a_3 \eta_3),
\nonumber\\
H_{3,4} &=&
\textstyle
(h_x+\frac{1}{2}\eta_1)
-(h-\frac{1}{2} a_1 \xi)
\mp(h_y-\frac{1}{2} a_2 \eta_2)
\pm(h_z+\frac{1}{2} a_3 \eta_3).
\label{fields}
\end{eqnarray}
Here dimensionless quantities $h = \Delta/S$, $h_{\alpha} =
d_{\alpha}E_{\alpha}/S$, $\Theta = k_{\mathrm{B}}T/S$,
$\beta=1/\Theta$,
\begin{eqnarray}
&&
a_1 = [(K_{13}+K_{14})-(J+K_{12})]/S,\qquad
a_2 = [(K_{13}-K_{14})-(J-K_{12})]/S,
\nonumber\\
&&
a_3 = [(K_{13}-K_{14})+(J-K_{12})]/S,\qquad
S = (K_{13}+K_{14})+(J+K_{12})
\label{dim-less}
\end{eqnarray}
and the symmetry properties of interaction constants are used. The
order parameters $\eta_1$, $\eta_2$, $\eta_3$ and the parameter
$\xi$ are determined from the set of equations
(\ref{set1}--\ref{fields}). Thermodynamically stable solutions are
those with the minimum values of the free energy. In absence of
the external field the solution $\eta_1 \neq 0$, $\xi \neq 0$,
$\eta_2=\eta_3=0$ corresponds to the ferroelectric phase in RS. In
this case $\langle S^z_1 \rangle = \langle S^z_2 \rangle$,
$\langle S^z_3 \rangle = \langle S^z_4 \rangle$ and the
four-sublattice model can be reduced to the Mitsui model. After
replacements $\frac{1}{2}\eta_1 \to \eta'$, $\frac{1}{2}\xi \to
\xi'$ and $F/(2N) \to F'/N'$ in equations
(\ref{set1}--\ref{fields}), one can obtain exactly the same
formulae as in that case.
Nonzero values $\eta_2 \neq 0$ or $\eta_3 \neq 0$ are induced in
paraphase by the corresponding components of the external field.
In the ferroelectric phase the parameters $\eta_2$ and $\eta_3$
are mutually connected. If one applies electric field along the
$b$-axis ($h_y \neq 0$, $\eta_2 \neq 0$) to the RS crystal in the
ferroelectric state ($\eta_1 \neq 0$), the third parameter
$\eta_3$ automatically becomes nonzero.

\section{Transverse field influence on polarization and
\protect\\
susceptibility}

The main advantage of our model is a possibility to describe the
dielectric properties and polarization of the RS crystal both
along and perpendicularly to the ferroelectric axis. There is also
an opportunity to consider the effects induced by the external
transverse field. Fig.~\ref{fig02} illustrates the possible dipole
orderings in some important cases when field and polarization are
parallel to the plane $XY$ ($ab$).

Components of tensor of the dielectric susceptibility
$\chi_{xx} = ({2d_x}/{\varepsilon_0 v_{\mathrm{c}}})
({\partial \eta_1}/{\partial E_x})$
and
$\chi_{yy} = ({2d_y}/{\varepsilon_0 v_{\mathrm{c}}})
({\partial \eta_2}/{\partial E_y})$
(where $v_{\mathrm{c}}$ is the unit cell volume) are determined
from the set of equations (\ref{set1}--\ref{fields}) by means of
implicit differentiation. Defining
$\chi_{\alpha\beta}=({2d_{\alpha}d_{\beta}}/{S\varepsilon_0
v_{\mathrm{c}}})\tilde{\chi}_{\alpha\beta}$,
we obtain e.g. for paraphase in presence of the field
$E_y$:
\begin{eqnarray}
\tilde{\chi}_{xx} &=&
\frac{R_1(8\Theta - \frac{1}{2} a_3 R_1) + \frac{1}{2} a_3 R_2^2}%
{(8\Theta - \frac{1}{2} R_1)(8\Theta - \frac{1}{2} a_3 R_1) - \frac{1}{4} a_3 R_2^2},
\\
\tilde{\chi}_{yy} &=&
\frac{R_1(8\Theta + \frac{1}{2} a_1 R_1) - \frac{1}{2} a_1 R_2^2}%
{(8\Theta + \frac{1}{2} a_1 R_1)(8\Theta + \frac{1}{2} a_2 R_1) - \frac{1}{4} a_1 a_2 R_2^2}.
\end{eqnarray}
Here
$R_1 = 4(1-\eta_2^2-\xi^2)$ and $R_2 = -8\eta_2\xi$.
In the case of the ferroelectric phase the parameters $\eta_1$ and
$\eta_3$ appear besides $\eta_2$ and $\xi$ in the expressions for
$\chi_{xx}$ and $\chi_{yy}$. Their temperature and field
dependences are determined from equations
(\ref{set1}--\ref{fields}).

As it follows from the analysis of behaviour of the free energy,
the second order type of the phase transitions remains unchanged
at $E_y \neq 0$. In such a case the transition temperatures can be
determined from the equation
\begin{equation}
\left(8\Theta - \frac{1}{2} R_1\right)
\left(8\Theta - \frac{1}{2} a_3 R_1\right)
- \frac{1}{4} a_3 R_2^2 = 0,
\end{equation}
which must be solved together with equations following from
(\ref{sz}) when $\eta_1 \to 0$, $\eta_3 \sim h_y \eta_1 \to 0$.

At small values of the transverse field
$R_1 = 4(1-\xi_0^2)+o[E_y^2]$,
%R_1 = 4(1-\xi_0^2)+{\scriptstyle\mathcal{O}}(E_y^2),
$R_2 \sim E_y^2$,
where $\xi_0$ is a solution of the equation
$\xi_0 = \tanh \left[\frac{1}{2}\beta\left(h- \frac{1}{2}a_1 \xi_0 \right)\right]$.
The critical temperatures shift under field proportionally to $E_y^2$.
Signs and absolute values of the shifts $\Delta T_{\mathrm{c}}$
depend on magnitudes and signs of the interaction parameters $a_2$
and $a_3$ as well as on the relation between them. From pure
geometric arguments the parameters $K_{12}$ and $J$ include
interactions between nearest and next nearest neighbours,
respectively. So one can expect that $K_{12} > J$; it results in
the inequality $a_2 > a_3$. When we use values
$a_1 = 0.284$ and $h = 0.32$
chosen to obtain the best fit for critical temperatures of the RS
crystal at zero field (in this case $S = 2280$~K),
the numerical analysis shows that at $a_3
\lesssim -0.25$ the ferroelectric region narrows under the field
$E_y$. The direct experimental verification is absent, but as some
evidence of such a possibility we can consider the results
obtained in \cite{Kal94,Fug03} by investigations of relaxation
phenomena in RS under the external transverse field.

\begin{figure}
\begin{center}
\begin{tabular}{ccc}
\includegraphics[width=0.4\textwidth]{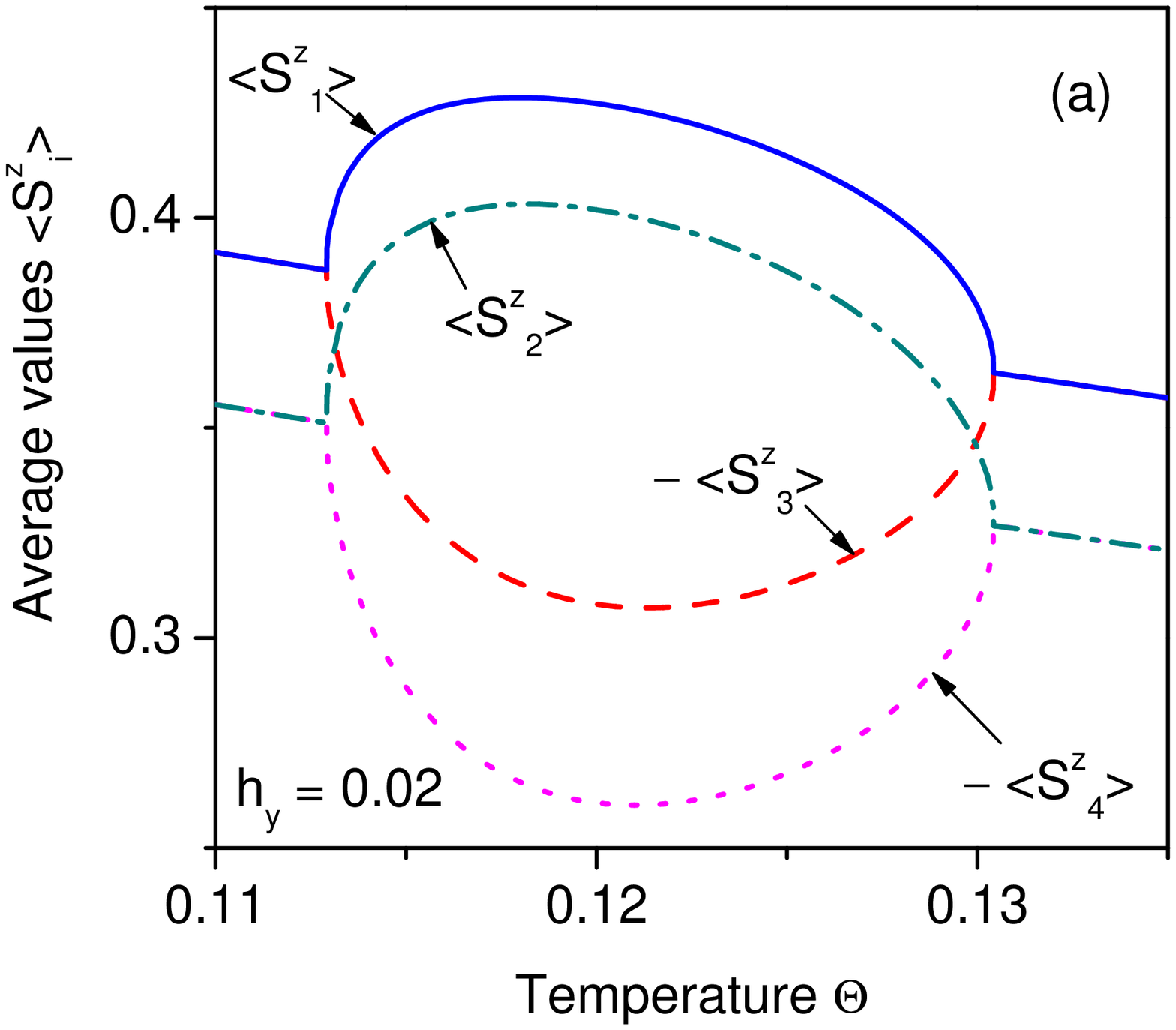}
&&
\includegraphics[width=0.4\textwidth]{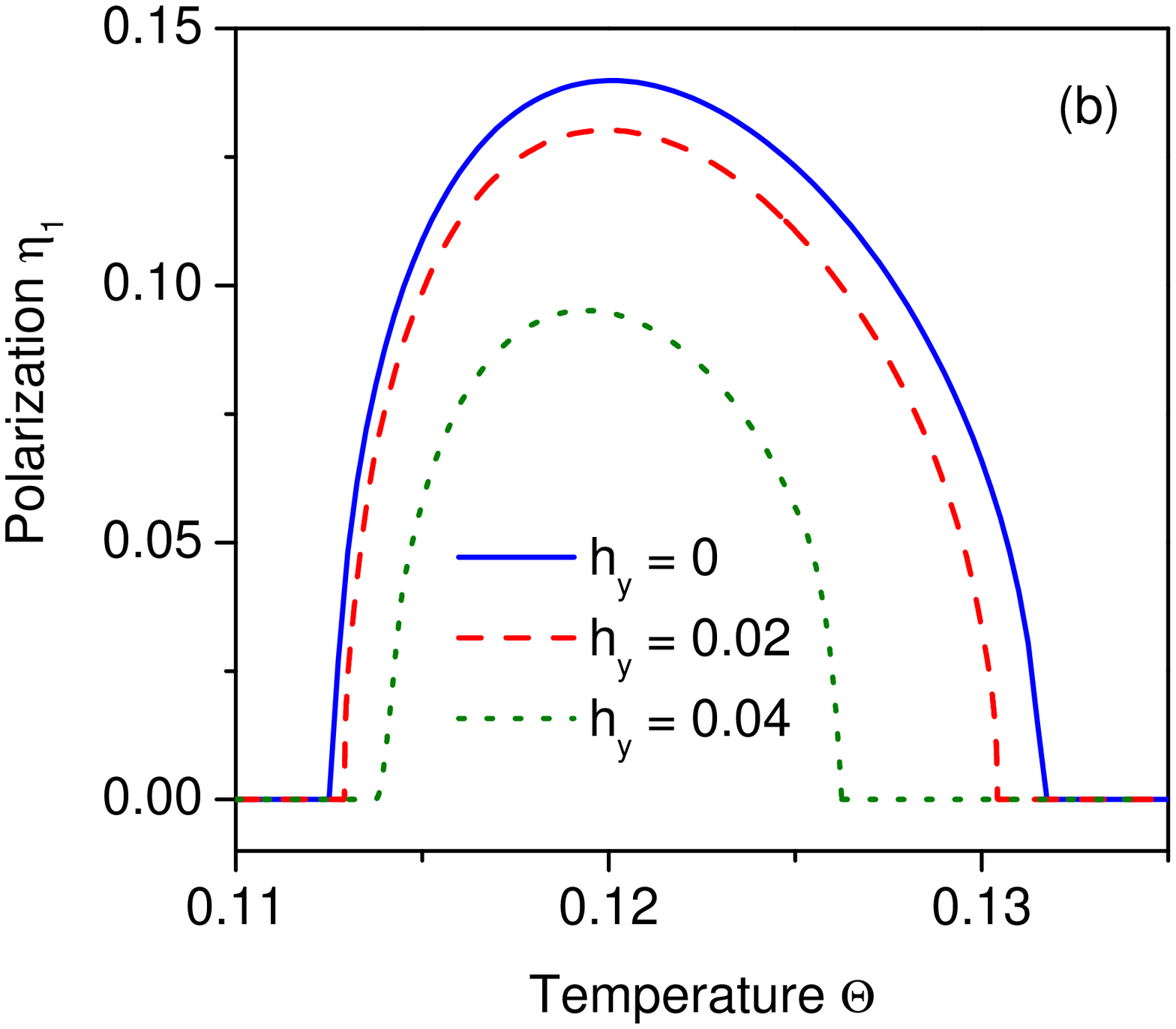}
%\\
%(a) && (b)
\\ [1.5ex]
\includegraphics[width=0.4\textwidth]{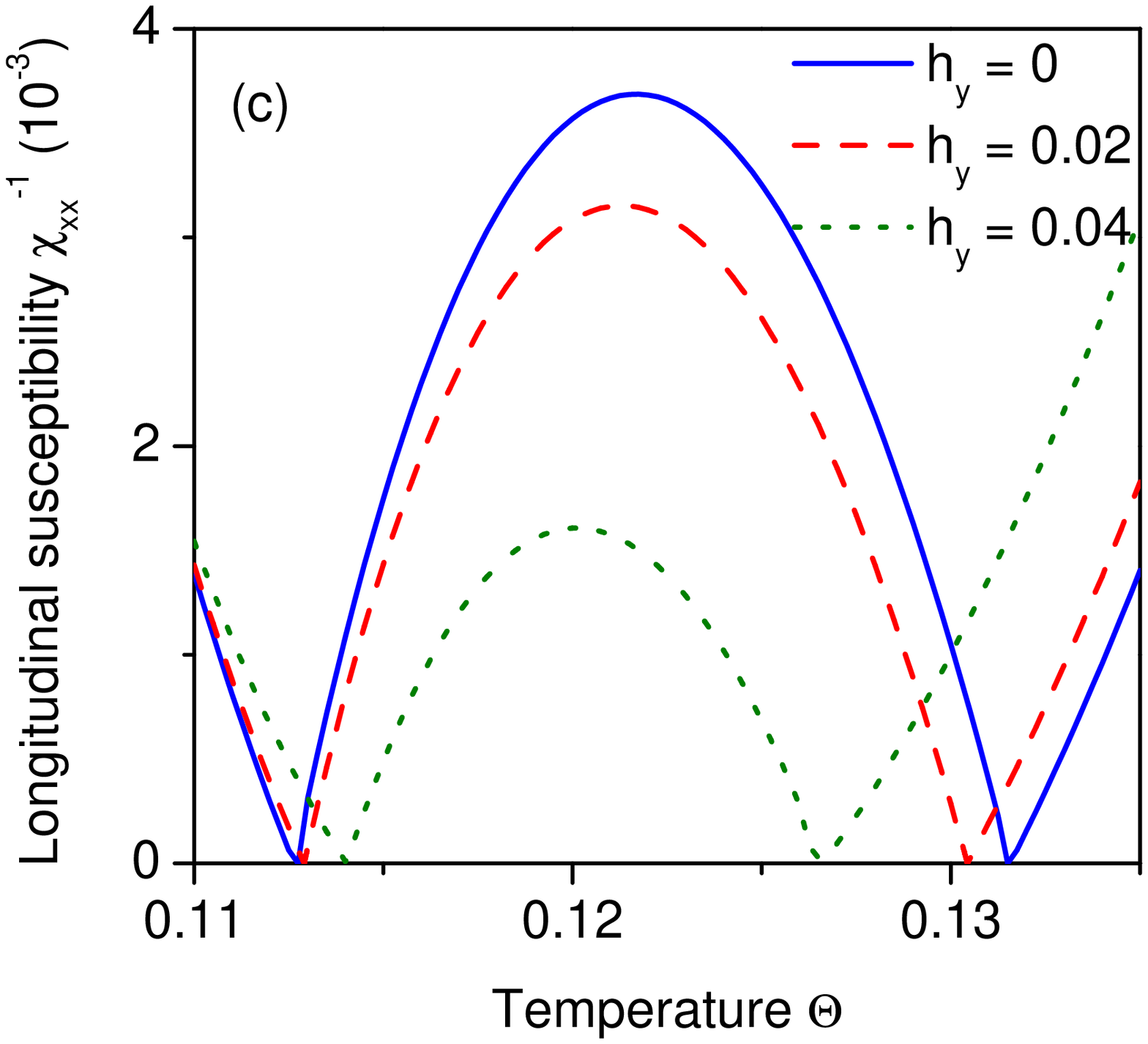}
&&
\includegraphics[width=0.43\textwidth]{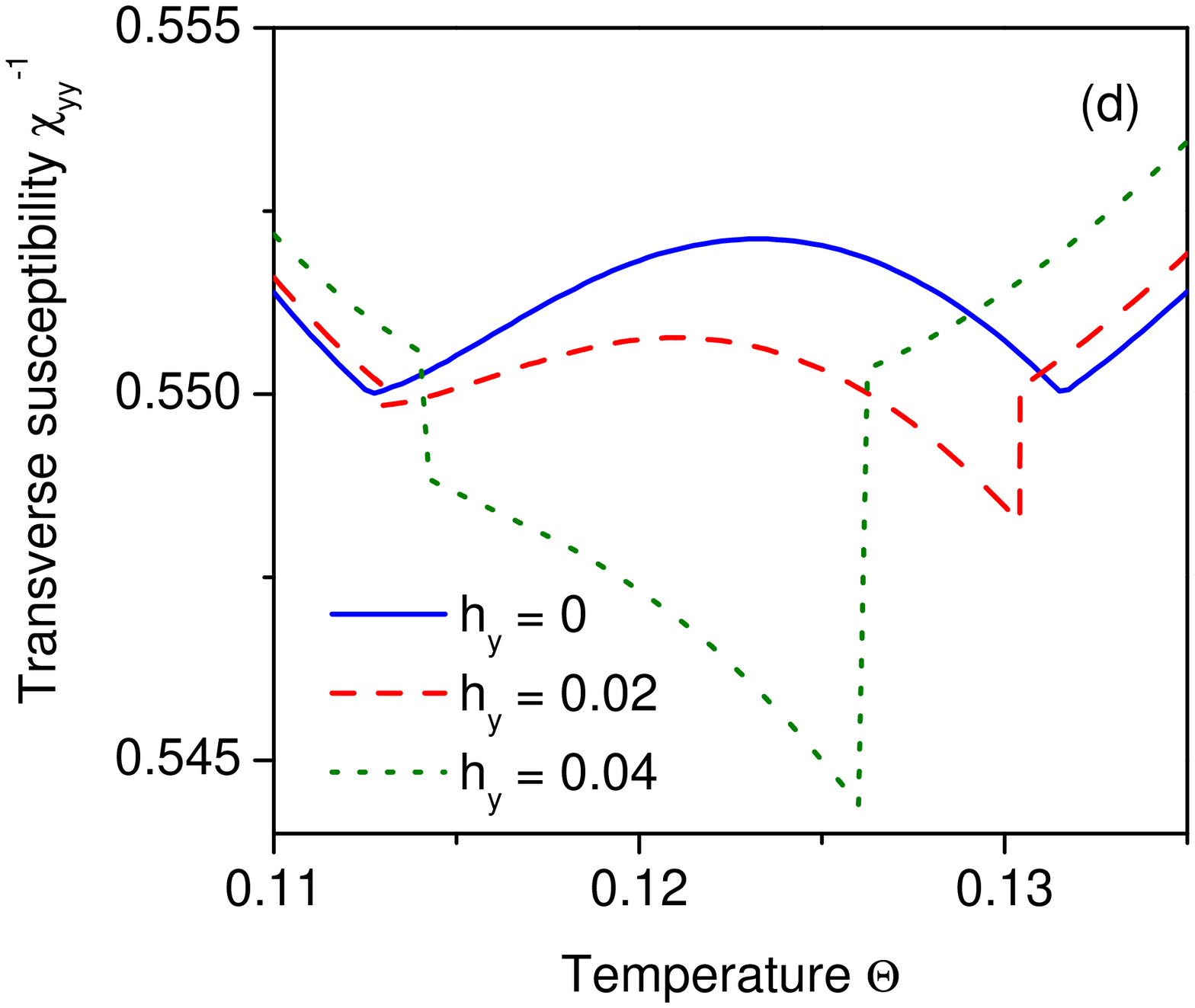}
%\\
%(c) && (d)
\end{tabular}
\end{center}
\caption{Temperature dependences of average pseudospin values (a),
the order parameter $\eta_1$
(proportional to the spontaneous polarization)
and longitudinal (c) and transverse (d) components
of the dielectric susceptibility
for different transverse fields at the following
values of parameters: $a_1 = 0.284$, $a_2 = 0.1$, $a_3 = -0.25$,
$h = 0.32$.}
\label{fig03}
\end{figure}

For illustration we give below numerical results for the
parameter values $a_2 = 0.1$ and $a_3 = -0.25$. Temperature
dependences of average pseudospin values in the cases depicted in
Fig.~\ref{fig02}(c,d) are presented in Fig.~\ref{fig03}(a). Pairs
$\langle S^z_1 \rangle$, $\langle S^z_3 \rangle$ and $\langle
S^z_2 \rangle$,  $\langle S^z_4 \rangle$ demonstrate a typical
``Mitsui-like'' behaviour but the transverse field splits their
values creating difference between pairs of sublattices even in
paraphase. Temperature dependences of the order parameter
$\eta_1$, describing spontaneous polarization in the RS crystal,
are shown in Fig.~\ref{fig03}(b) for different values of the
transverse field. One can see that such a field not only narrows
the temperature range of the ferroelectric phase but can also
suppress spontaneous polarization.

Temperature behaviour of components of the dielectric
susceptibility is shown in Fig.~\ref{fig03}(c,d). The inverse
susceptibility $\chi_{xx}^{-1}$ goes to zero in the phase
transition points both at $E_{y}=0$ and $E_{y} \neq 0$
(Fig.~\ref{fig03}(c)). This fact confirms that the phase
transitions are of the second order.
The transverse component $\chi_{yy}^{-1}$ has jumps in the
transition points at $E_{y} \neq 0$. Their values are proportional
to the second power of the field magnitude (Fig.~\ref{fig03}(d)).
These jumps closely resemble behaviour of the transverse
susceptibility of the glycinium phosphite crystals in the
transverse field \cite{Sta04}.

Let us make some numerical estimates taking into account the
obtained results and using the experimental data for
$\varepsilon_a$ and $\varepsilon_b$ components and for
$P_{\mathrm{S}}$ at $E_y = 0$. The dipole moment component $d_x$
can be determined using the maximal value of $P_{\mathrm{S}}$ in
the ferroelectric phase ($P_{\mathrm{S}}|_{\mathrm{max}}=0.25
\times 10^{-2}$~C/m$^2$ \cite{Jon62}). From the relation
$P_{\mathrm{S}}=(2d_x/v_{\mathrm{c}})\eta_1$, at
$\eta_1|_{\mathrm{max}}=0.14$ and $v_{\mathrm{c}}=1.04 \times
10^{-21}$~cm$^{-3}$, we obtain $d_x = 9.26 \times 10^{-30}$~C\,m.
Respectively, for susceptibility along the $a$-axis we have
$\chi_{xx}=0.60 \tilde{\chi}_{xx}$, and at
$\tilde{\chi}_{xx}^{-1}|_{\mathrm{max}} = 3.7 \times 10^{-3}$
it results in $\chi_{xx}|_{\mathrm{min}} \simeq 160$ (such a value
lies inside the experimentally observed range of $\chi_{xx}$ for
ferroelectric phase, see review in \cite{Jon62,Smo71}).

Estimate for $d_y$ component can be obtained using the relation
$\varepsilon_{yy}=1+(2{d_y^2}/{S\varepsilon_0
v_{\mathrm{c}}})\tilde{\chi}_{yy}$. In the ferroelectric phase
region $\varepsilon_{yy}\equiv\varepsilon_{b}\approx 10$ (see
\cite{Smo71}; the old experimental data show the smooth
temperature dependence of $\varepsilon_{b}$ in this region). At
$\tilde{\chi}_{yy}^{-1} = 0.552$ we have $d_y= 17.3 \times
10^{-30}$~C\,m. So the $Y$-component of a dipole moment, connected
with reordering, is nearly twice as large as the one along the
ferroelectric $X$-axis.

Let us notice that in this case the field $E_y = 18$~MV/m
corresponds to the value $h_y = 0.01$; the shift of
$T_{\mathrm{c}1}$ is $\Delta T_{\mathrm{c}1} \approx
0.06$~K at that field. It means that at the fields $E_y \approx
1$~MV/m the effect will be practically undetectable. The relative
change of the susceptibility $\chi_{yy}$ with temperature in the
ferroelectric phase region is also small
\linebreak%
($\approx 0.5\%$).
However, the results of numerical estimates can change at another choice
of the parameter values $a_2$ and $a_3$, so the field effect can be much
stronger. As a certain argument which points to such a
possibility, we can consider the fact, that in the GPI crystal
(where the effect is caused, as in RS, by the zig-zag
geometry of the local dipole moment arrangement) the change of
$T_{\mathrm{c}}$ is $\Delta T_{\mathrm{c}} \approx 0.05$~K at the
transverse field $E_c \approx 1$~MV/m. It is obvious that one can
check the possibility of such a noticeable effect in RS only
by direct experimental investigations.

\section{Conclusions}

For description of phase transitions and dielectric properties of
the Rochelle salt crystal we propose the four-sublattice
pseudospin model developed as a generalization of the well-known
Mitsui model. The introduced model takes into account spatial
orientations of effective dipoles, which are related to the atomic
groups in the unit cell and are responsible for the spontaneous
polarization.

The model allows to investigate the temperature behaviour of both
longitudinal and transverse components of dielectric
susceptibility as well as to consider the influence of the
transverse electric field $\vec{E}\parallel\vec{b}$. At the
certain relations between the model parameter values the increase
of this field can lead to the approaching of the lower and higher
Curie points one to another. Our theory predicts suppression of
the spontaneous polarization $P_{\mathrm{S}}$ under the field
$E_y$. The effect is similar to the one observed in
\cite{Kal94,Fug03}. The changes in the transverse susceptibility
$\chi_{yy}$ in the transition points are similar to the phenomena
detected in the GPI crystal \cite{Sta04}. As it follows from our
consideration, there should exist jumps in $\chi_{yy}$ which
increase proportionally to $E_y^2$.
However, performed here numerical estimates indicate that these
effects could have small magnitudes. Only future experimental
studies can give a final answer.

The ideas forming the basis of the four-sublattice model can be
used for description of the mixed system RS$_{1-x}$--ARS$_{x}$,
where the significant changes in temperature behaviour of the
longitudinal and transverse dielectric susceptibilities are
observed at increase of the concentration $x$ of ammonia groups
and in the high concentration region ($0.89<x<1$) a polar phase
appears with polarization along the $b$-axis \cite{Sch00,Kik03}.

\section*{Acknowledgments}

Oleh Velychko is deeply indebted to Prof.~R.~Nozaki,
Dr.~P.~Lunkenheimer, Dr.~S.~Kamba, Dr.~J.~Kulda and Dr.~B.~Fugiel
for imprints of their articles and expresses special thanks to the
old friends Dr.~Kyrylo Tabunshchyk and Dr.~Oleh Danyliv for
sending him the rest of literature. This work would't be
finished without their kind help.

\end{document}